\documentclass[aps,prl,twocolumn,groupedaddress,preprintnumbers,nofootinbib,balancelastpage]{revtex4-1}

\usepackage{amssymb}
\usepackage{graphicx}
\usepackage{amsmath}
\usepackage{hyperref}
\usepackage{xspace}

\newcommand{\as}{\alpha_s}

\newcommand{\zcut}{z_\text{cut}}

\DeclareRobustCommand{\Tab}[1]{Table~\ref{#1}}

\DeclareRobustCommand{\Fig}[1]{Fig.~\ref{#1}}

\DeclareRobustCommand{\Eq}[1]{(\ref{#1})}
\DeclareRobustCommand{\Eqs}[2]{(\ref{#1}) and (\ref{#2})}
\DeclareRobustCommand{\Ref}[1]{\cite{#1}}
\DeclareRobustCommand{\Refs}[1]{\cite{#1}}

\newcommand{\pythia}[1]{\textsc{Pythia\xspace #1}}

\newcommand{\fastjet}[1]{\textsc{FastJet\xspace #1}}

\newcommand{\herwigpp}[1]{\textsc{Herwig++\xspace #1}}
\newcommand{\vincia}[1]{\textsc{Vincia\xspace #1}}
\newcommand{\sherpa}[1]{\textsc{Sherpa\xspace #1}}

\usepackage{color}
\definecolor{darkblue}{rgb}{0,0,0.5}
\definecolor{darkred}{rgb}{0.5,0,0}
\definecolor{darkgreen}{rgb}{0,0.5,0}

\newcommand{\df}{\mathrm{d}}
\newcommand{\rg}{r_g}
\newcommand{\zg}{z_g}
\newcommand{\amax}{a_2}
\newcommand{\amin}{a_1}

\begin{document}

\preprint{MIT-CTP 4638}

\title{Sudakov Safety in Perturbative QCD}

\author{Andrew J.~Larkoski}
\email{larkoski@mit.edu}

\affiliation{Center for Theoretical Physics, Massachusetts Institute of Technology, Cambridge, MA 02139, USA}

\author{Simone Marzani}
\email{smarzani@mit.edu}

\affiliation{Center for Theoretical Physics, Massachusetts Institute of Technology, Cambridge, MA 02139, USA}

\author{Jesse Thaler}
\email{jthaler@mit.edu}

\affiliation{Center for Theoretical Physics, Massachusetts Institute of Technology, Cambridge, MA 02139, USA}

\date{\today}

\begin{abstract}
Traditional calculations in perturbative quantum chromodynamics (pQCD) are based on an order-by-order expansion in the strong coupling $\alpha_s$.  Observables that are calculable in this way are known as ``safe''.  Recently, a class of unsafe observables was discovered that do not have a valid $\alpha_s$ expansion but are nevertheless calculable in pQCD using all-orders resummation.  These observables are called ``Sudakov safe'' since singularities at each $\alpha_s$ order are regulated by an all-orders Sudakov form factor.  In this paper, we give a concrete definition of Sudakov safety based on conditional probability distributions, and we study a one-parameter family of momentum sharing observables that interpolate between the safe and unsafe regimes.  The boundary between these regimes is particularly interesting, as the resulting distribution can be understood as the ultraviolet fixed point of a generalized fragmentation function, yielding a leading behavior that is independent of $\alpha_s$.
\end{abstract}

\pacs{}
\maketitle

Infrared and collinear (IRC) safety has long been a guiding principle for determining which observables are calculable using perturbative quantum chromodynamics (pQCD) \cite{Sterman:1977wj,Ellis:1991qj}. IRC safe observables are insensitive to arbitrarily soft gluon emissions and arbitrarily collinear parton splittings.  This property ensures that perturbative singularities cancel between real and virtual emissions, leading to finite cross sections order-by-order in the strong coupling $\alpha_s$.  At the Large Hadron Collider (LHC), IRC safe jet algorithms like anti-$k_T$ \cite{anti-kt} play a key role in almost every analysis, and many jet-related cross sections have been calculated to next-to-leading and even next-to-next-to-leading order \cite{Ridder:2013mf, Currie:2013dwa,Boughezal:2013uia,Chen:2014gva}.  Of course, there are observables relevant for collider physics that are not IRC safe, though one can often use non-perturbative objects---like parton distribution functions, fragmentation functions (FFs), and their generalizations \cite{Krohn:2012fg,Waalewijn:2012sv,Chang:2013rca,Chang:2013iba}---to absorb singularities and restore calculational control.

In this paper, we show how to extend the calculational power of pQCD into the IRC unsafe regime using purely perturbative techniques.  We study a class of unsafe observables that are not defined at any fixed order in $\alpha_s$, yet nevertheless have finite cross sections when all-orders effects are included.  These observables are known in the literature as ``Sudakov safe'' \cite{Larkoski:2013paa}, since a perturbative Sudakov form factor \cite{Sudakov:1954sw} naturally (and exponentially) regulates real and virtual infrared (IR) divergences.  To date, however, the study of Sudakov safe observables has been limited to specific examples.  Here, we achieve a deeper understanding of these observables by providing a concrete definition of Sudakov safety based on conditional probabilities.   The techniques in this paper apply to any perturbative quantum field theory, but we focus on pQCD to highlight an example of direct relevance to jet physics at the LHC.

Because Sudakov safe observables are not defined at any fixed perturbative order, they in general have non-analytic dependence on $\alpha_s$.  Examples in the literature include observables with an apparent expansion in $\sqrt{\alpha_s}$ \cite{Larkoski:2013paa} and observables which are independent of $\alpha_s$ at sufficiently high energies \cite{Larkoski:2014wba,Larkoski:2014bia}.  As a case study, we consider a one-parameter family of momentum sharing observables $z_g$ based on ``soft drop declustering'' \cite{Larkoski:2014wba}, which already appears in many jet substructure studies, e.g.~\cite{BDRS,taggersRES,taggersNLO}.  This family not only interpolates between the above two Sudakov-safe behaviors but also includes an IRC-safe regime.  We explain how the boundary between the safe and unsafe regimes can be understood using the more familiar language of (generalized) FFs; the renormalization group (RG) evolution of the FF has an ultraviolet (UV) fixed point, suggesting an extended definition of IRC safety.

To begin our general discussion of Sudakov safety, consider an IRC unsafe observable $u$ and a companion IRC safe observable $s$.  The observable $s$ is chosen such that its measured value regulates all singularities of $u$.  That is, even though the probability of measuring $u$,
\begin{equation}
p(u) = \frac{1}{\sigma} \frac{\df \sigma}{\df u},
\end{equation}
is ill-defined at any fixed perturbative order, the probability of measuring $u$ given $s$, $p(u|s)$, is finite at all perturbative orders, except possibly at isolated values of $s$; e.g., $s=0$.  Given this companion observable $s$, we want to know whether $p(u)$ can be calculated from pQCD.

Because $s$ is IRC safe, $p(s)$ is well-defined at all perturbative orders (although resummation may be required to regulate isolated singularities, see below).  
This allows us to define the joint probability distribution
\begin{equation}\label{eq:cond_prob}
p(s,u) = p(s)\, p(u|s) ,
\end{equation}
which is also finite at all perturbative orders, except possibly at isolated values of $s$.  To calculate $p(u)$, we can simply marginalize over $s$:
\begin{equation}
\label{eq:sudsafeone}
p(u) = \int \df s\, p(s) \, p(u|s) \,.
\end{equation}
If $p(s)$ regulates all (isolated) singularities of $p(u|s)$, thus ensuring that the above integral is finite, then we deem $u$ to be Sudakov safe.
In the case that one IRC safe observable is insufficient to regulate all singularities in $u$, we can measure a vector of IRC safe observables ${\bf s}=\{s_1,\dotsc,s_n\}$.  This gives a more general definition of Sudakov safety:
\begin{equation}\label{eq:sudsafedef}
p(u)= \int \df^n{\bf s} \, p({\bf s})\, p(u|{\bf s}) \,.
\end{equation}

All previous examples of Sudakov safety fall in the category of \Eq{eq:sudsafeone} above where only a single IRC safe compansion $s$ was required.   In \Ref{Larkoski:2014wba}, the energy loss distribution from soft drop grooming was defined precisely as in \Eq{eq:sudsafeone}, where $u$ was the factional energy loss $\Delta_E$ and $s$ was the groomed jet radius $\rg$ (see below). In \Ref{Larkoski:2013paa}, ratio observables $r = a/b$ were originally defined in terms of a double-differential cross section \cite{Larkoski:2014tva,Procura:2014cba} as
\begin{equation}
\label{eq:ratio_doublediff}
p(r) = \int \df a \, \df b \, p(a,b) \, \delta\left(r - \frac{a}{b} \right),
\end{equation}
where $a$ and $b$ are IRC safe but $r$ is not, because there are singularities at $b=0$ at every finite perturbative order, leading to a divide-by-zero issue for $r$.  Integrating over $a$ and using the definition of conditional probability \Eq{eq:cond_prob}, we can write \Eq{eq:ratio_doublediff} as
\begin{equation}
\label{eq:ratio_conditional}
p(r) = \int \df b \, p(b) \, p(r|b) \,,
\end{equation}
and $r$ is Sudakov safe because $p(b)$ has an all-orders Sudakov form factor that renders $p(r)$ finite.

It should be stressed that the definition of Sudakov safety in \Eq{eq:sudsafedef} is not vacuous and it does not save all IRC unsafe observables.  As a counterexample, consider particle multiplicity;  because perturbation theory allows an arbitrary number of soft or collinear emissions, one would need to measure an infinite number of IRC safe observables to regulate all singularities to all orders.  Also, it should be stressed that just because an observable is Sudakov safe, that does not imply that non-perturbative aspects of QCD are automatically suppressed.  While a detailed discussion is beyond the scope of this paper, both \Refs{Larkoski:2014wba,Larkoski:2013paa} include an estimate of non-perturbative effects, which are analogous to power corrections and underlying event corrections familiar from the IRC safe case.  In some cases, these corrections are known to scale away as a (fractional) inverse power of the collision energy.

Crucially, one needs some kind of all-orders information to obtain finite distributions for $p(u)$.  If a fixed-order expansion of $p(s)$ and $p(u|s)$ were sufficient, then $p(u)$ would have a series expansion in $\alpha_s$, contradicting the assumption that $u$ is IRC unsafe. In this paper, we use logarithmic resummation to capture all-orders information about $p(s)$, which regulates isolated singularities at $s=0$ to ensure the integral in \Eq{eq:sudsafeone} is finite.  In all cases we have encountered, a finite $p(u|s)$ with a resummed $p(s)$ is sufficient to calculate $p(u)$, though this may not be the case generally.

Unlike IRC safe distributions which have a unique $\alpha_s$ expansion, the formal perturbative accuracy of a Sudakov safe distribution is potentially ambiguous. First, there are different choices for $s$ that can regulate the singularities in $u$.  This is analogous to the choice of evolution variables in a parton shower, as each choice gives a finite (albeit different) answer at a given perturbative accuracy.  Second, the probability distributions $p(s)$ and $p(u|s)$ can be calculated to different formal accuracies.  Below we use leading logarithmic resummation for $p(s)$, but only work to lowest order in $\alpha_s$ for $p(u|s)$.   Thus, when discussing the accuracy of $p(u)$, one must specify the choice of $s$ and the accuracy of $p(s)$ and $p(u|s)$ separately.  We stress, however, that the accuracy of both objects is systematically improvable.

We now study an instructive example that demonstrates the complementarity of Sudakov safety and IRC safety.  This example is based on soft drop declustering \cite{Larkoski:2014wba}, which we briefly review.  Consider a jet clustered with the Cambridge-Aachen (C/A) algorithm~\cite{Wobisch:1998wt, Dokshitzer:1997in} with jet radius $R_0$.  One can decluster through the jet's branching history, grooming away the softer branch until one finds a branch that satisfies the condition
\begin{equation}\label{eq:sdcrit}
\frac{\min \left( p_{T1}, p_{T2}\right)}{p_{T1} + p_{T2}} > z_\text{cut}\left( \frac{R_{12}}{R_0}\right)^\beta,
\end{equation}
where $1$ and $2$ denote the branches at that step in the clustering, $p_{Ti}$ are the corresponding transverse momenta, and $R_{12}$ is their rapidity-azimuth separation.  The kinematics of this branch defines the groomed jet radius $\rg$ and the groomed momentum sharing $\zg$,
\begin{equation}
\rg = \frac{R_{12}}{R_0}, \qquad \zg = \frac{\min \left( p_{T1}, p_{T2}\right)}{p_{T1} + p_{T2}};
\end{equation}
$\rg$ is IRC safe and its distribution was studied in \Ref{Larkoski:2014wba}.

Our observable of interest is $\zg$, and the angular exponent $\beta$ determines whether or not $\zg$ is IRC safe.  For $\beta < 0$, $\zg$ is IRC safe, because $\zg>z_\text{cut}$ for any branch that passes \Eq{eq:sdcrit}; if this condition is never satisfied, the jet is simply removed from the analysis.  For $\beta > 0$, $\zg$ is IRC unsafe, since measuring $\zg$ does not regulate collinear singularities.  The boundary case $\beta = 0$ corresponds to the (modified) mass drop tagger \cite{BDRS,taggersRES,taggersNLO} which also has collinear divergences, but we will show that it actually satisfies an extended version of IRC safety.

In our calculations, we work to lowest non-trivial order to illustrate the physics, though we provide supplemental materials for the interested reader that include higher-order (and non-perturbative) effects.  We take the parameter $z_\text{cut}$ to be small, but large enough that $\log z_\text{cut}$ terms need not be resummed, with a benchmark of $z_\text{cut} \simeq 0.1$.

  \begin{figure}[t]
  \includegraphics[height=5cm]{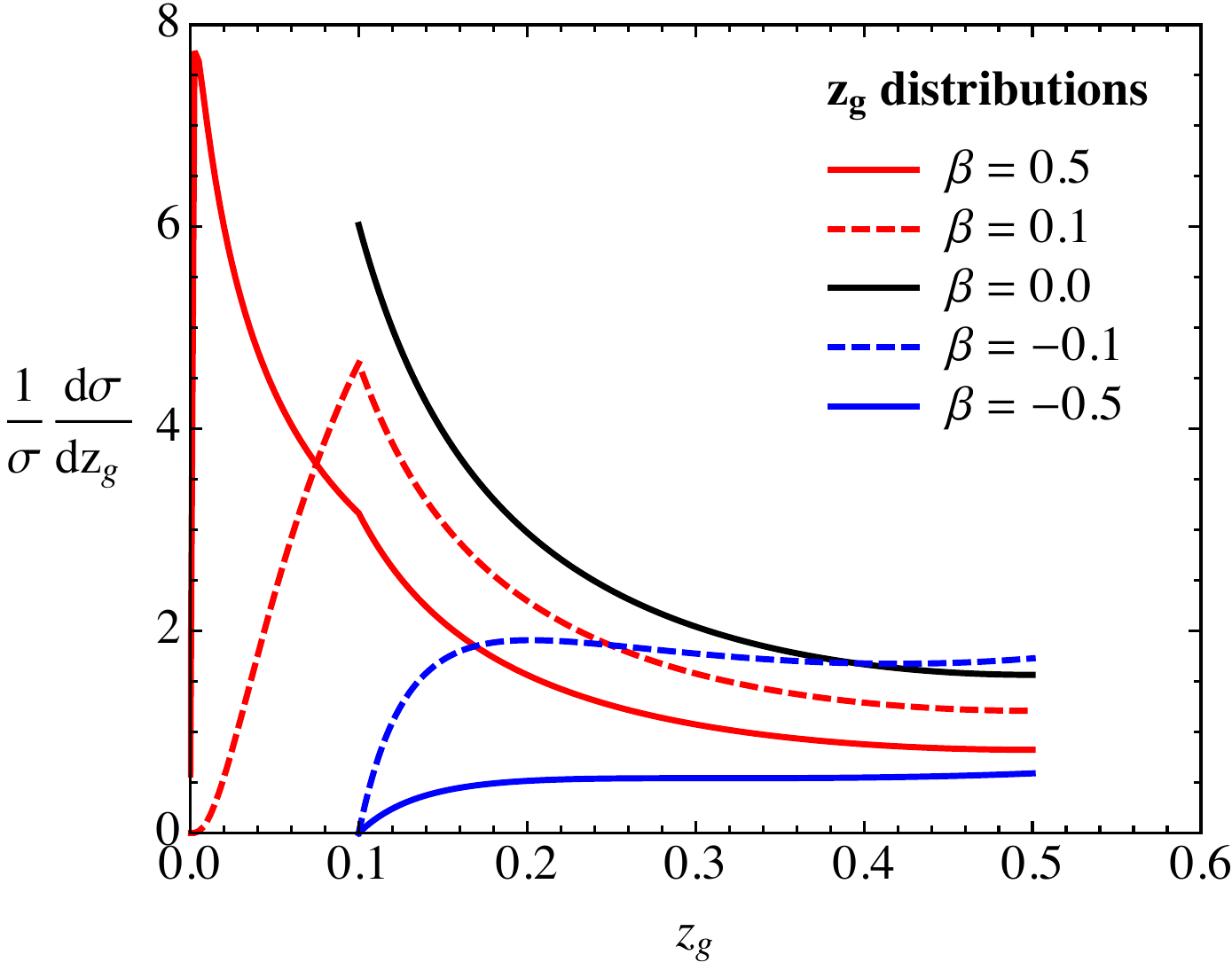}
 \caption{Distributions of $\zg$ for various $\beta$ values, obtained from \Eq{eq:zfeq} at fixed $\as=0.1$ and $\zcut = 0.1$. 
}
 \label{fig:zf_an_dists}
 \end{figure}

We now use the strategy in \Eq{eq:sudsafeone} to calculate the momentum sharing $\zg$ for all values of $\beta$, using the groomed radius $\rg$ to regulate collinear singularities:
\begin{equation}\label{eq:zfeq}
p(\zg)=\frac{1}{\sigma}\frac{\df \sigma}{\df \zg} = \int \df \rg \, p(\rg) \, p(\zg|\rg) \,.
\end{equation}
We use all-orders resummation to determine $p(\rg)$ and regulate the isolated $r_g = 0$ singularity.  This has been carried out to next-to-leading-logarithmic accuracy in \Ref{Larkoski:2014wba}.  Here, it is sufficient to consider the fixed-coupling limit:
\begin{equation} \label{eq:resum_r}
p(r_g) =  \frac{\df}{\df \rg} \exp \left[ -\frac{2 \as C_i}{\pi} \int_{\rg}^1\frac{\df \theta}{\theta}\int_0^1 \df z \, P_i(z)\,\Theta_\text{cut} \right] \,,
\end{equation}
where $C_i$ is the color factor of the jet, $P_i(z)$ is the appropriate splitting function (summed over final states), and the phase space cut is
\begin{align}\label{eq:theta_cut}
\Theta_\text{cut}&= \Theta(1/2-z)\Theta\left(z-z_\text{cut}\theta^\beta\right)\nonumber \\
&\qquad+ \Theta\left(z-1/2\right)\Theta\left((1-z)-z_\text{cut}\theta^\beta\right)\,.
\end{align}
The exponential part of \Eq{eq:resum_r} is the $r_g$ Sudakov form factor, where $\Theta_\text{cut}$ defines the no-emission criteria. To calculate $p(\zg|\rg)$, note that $\zg$ is defined by a single emission in the jet.  For small $R_0$, the lowest-order matrix element is well-approximated by a $1\to 2$ splitting function:
\begin{equation}
p(\zg|\rg ) =  \frac{\overline{P}_i(\zg)}{\int_{z_\text{cut}r_g^\beta}^{1/2} \df z \, \overline{P}_i(z)} \Theta(\zg-\zcut \rg^\beta)\,,
\end{equation}
where $0 < \zg < 1/2$ and we have introduced the notation
\begin{equation}
\overline{P}_i(z)=P_i(z)+P_i(1-z).
\end{equation}
In the double-logarithmic limit, we simply have $\overline{P}_i(z)= 1/z$, allowing an explicit evaluation of \Eq{eq:zfeq}:
\begin{align} \label{eq:res_zf}
p(\zg) &=\sqrt{\tfrac{\as C_i}{\beta}}  \exp \left[{\tfrac{\as C_i}{\pi \beta}\log^2 \tfrac{1}{2z_\text{cut}}} \right] \overline{P}_i(\zg) \\
& \quad\times \left( \text{erf} \left[ \sqrt{\tfrac{\as C_i}{\pi \beta} } \log \tfrac{1}{\amin} \right] -\text{erf} \left[ \sqrt{\tfrac{\as C_i}{\pi \beta} }\log \tfrac{1}{\amax}\right] \right) \nonumber,
\end{align}
where
\begin{align}
\beta \geq 0: & \quad  \amin = 0, &&  \amax = \min\left[2\zcut,2 \zg \right], 
 \\
\beta < 0: & \quad   \amin = 2 \zg, && \amax = 2\zcut.
\end{align}
Because \Eq{eq:res_zf} is finite, we see that $\zg$ is at least Sudakov safe for all $\beta$.  Distributions of $\zg$ calculated with \Eq{eq:zfeq} at fixed $\as$ are shown in \Fig{fig:zf_an_dists}.

By expanding $p(\zg)$ in small $\alpha_s$, we can better understand the difference between IRC-safe and Sudakov-safe behavior.  For $\beta<0$, $\zg$ is IRC safe, so $\zg$ should have a well-defined expansion in $\alpha_s$.  To the accuracy calculated, \Eq{eq:zfeq} is fully valid to ${\cal O}(\alpha_s)$ in the collinear limit, and the expansion of \Eq{eq:zfeq} yields the expected IRC safe result:
\begin{align}
\beta < 0: \quad p(\zg) &= \frac{2\alpha_s C_i}{\pi |\beta| } \, \overline{P}_i(\zg) \log\frac{\zg}{z_\text{cut}} \Theta(\zg -\zcut) \nonumber \\
&\quad +{\cal O}(\alpha_s^2) \,.
\end{align}
For $\beta > 0$, $\zg$ is only Sudakov safe and its distribution should not have a valid Taylor series in $\alpha_s$.  Indeed, for $\beta > 0$, the distribution has the expansion
\begin{align}
\beta > 0: &\quad p(\zg) = \sqrt{\frac{\alpha_s \, C_i}{ \beta}}\, \overline{P}_i(\zg)+{\cal O}\left(\alpha_s\right) , 
\end{align}
and the presence of $\sqrt{\alpha_s}$ implies non-analytic dependence on $\alpha_s$.  To ${\cal O}(\sqrt{\alpha_s})$, the only phase space constraint is $0<\zg <1/2$, and the kink visible in \Fig{fig:zf_an_dists} at $\zg =\zcut$ first appears at ${\cal O}(\alpha_s)$.  Finally, for the boundary case $\beta = 0$, $p(\zg|\rg )$ is independent of $\rg$ (in the fixed-coupling approximation), and \Eq{eq:res_zf} is independent of $\alpha_s$:
\begin{equation}
\label{eq:beta_zero_pzg}
\beta = 0: \quad p(\zg) = \frac{\overline{P}_i(\zg)}{\int_{z_\text{cut}}^{1/2} dz \, \overline{P}_i(z)}\Theta(\zg-\zcut)\,.
\end{equation}
We will later show that the $\beta = 0$ case does have a valid perturbative expansion in $\alpha_s$, despite being $\alpha_s$-independent at lowest order.  The behavior of $\zg$ for different $\beta$ values is summarized in \Tab{Tab:behavior}.

\begin{table}
\begin{tabular}{cccc}
\hline \hline
\phantom{aaaaaai}& Safety& Divergences & Expansion \\
\hline 
$\beta < 0$ & IRC & None & $\alpha_s^n$\\
$\beta = 0$ & \phantom{a}IRC via FF\phantom{a} & Collinear~Only & $\alpha_s^{n-1}$\\
$\beta > 0$ & Sudakov & \phantom{a}Collinear \& Soft-Coll.\phantom{a} & $\alpha_s^{n/2}$\\
\hline \hline
\end{tabular}
\caption{As $\beta$ is adjusted, $p(z_g)$ interpolates between IRC-safe and two Sudakov-safe behaviors, related to the divergences in $z_g$.  Here, $n \geq 1$ ranges over positive integers.
}
\label{Tab:behavior}
\end{table}

The $\beta = 0$ distribution of $\zg$ is fascinating (and simpler than previous $\alpha_s$-independent examples \cite{Larkoski:2014wba,Larkoski:2014bia}).  Because $\zg$ only has collinear divergences, we can understand $p(\zg)$ in a different and illuminating way using FFs.   As is well known, FFs absorb collinear divergences in final-state parton evolution, and we can introduce a generalized FF, $F(\zg)$, to play the same role for $\zg$.  In the standard case, FFs are non-perturbative objects with perturbative RG evolution.  In the $\zg$ case, $F(\zg)$ is still a non-perturbative object, but it has a perturbative UV fixed point, becoming independent of IR boundary conditions at sufficiently high energies.  

At Born level, the jet has a single parton, so $\zg$ is undefined.  We can, however, define $F(\zg)$ to be the one-prong $\zg$ distribution, such that $F(\zg)$ acts like a non-trivial measurement function that is independent of the kinematics.  Working to ${\cal O}(\alpha_s)$ in the collinear limit,
\begin{align}\label{eq:ff1}
p(\zg) &= F(\zg)\nonumber +\frac{\alpha_s C_i}{\pi}\int_0^1 \frac{d\theta}{\theta} \\
&
\times \left( \overline{P}_i(\zg) \Theta(\zg-\zcut) - F(\zg) \int_{z_\text{cut}}^{1/2}dz\, \overline{P}_i(z)\right) \nonumber \\
&
+{\cal O}(\alpha_s^2) \,.
\end{align}
There are two terms at ${\cal O}(\alpha_s)$.  The first term accounts for the resolved case where the jet is composed of two prongs from a $1 \to 2$ splitting.   The second term corresponds to additional one-prong configurations (with the same $F(\zg)$ measurement function as the Born case), arising either because the other prong has been removed by soft drop grooming or from one-prong virtual corrections.

For a general $F(\zg)$, \Eq{eq:ff1} is manifestly collinearly divergent because of the $\theta$ integral, and $F(\zg)$ must be renormalized.  But there is a unique choice of $F(\zg)$ for which collinear divergences are absent (at this order), without requiring renormalization:
\begin{align}\label{eq:f_uv}
F_\text{UV}(\zg)=\frac{\overline P_i(\zg)}{\int_{z_\text{cut}}^{1/2} dz \, \overline P_i(z)} \Theta(\zg-\zcut) \,.
\end{align}
Plugging this into \Eq{eq:ff1}, the $\mathcal{O}(\alpha_s)$ term vanishes, and we recover precisely the distribution in \Eq{eq:beta_zero_pzg}.

In this way, $\zg$ at $\beta = 0$ exhibits an extended version of IRC safety, where a non-trivial (and finite) measurement function is introduced in a region of phase space where the measurement would be otherwise undefined.  Similar measurement functions appeared (without discussion) in the early days of jet physics \cite{Pi:1978hu,Kramer:1978uj}, where symmetries determined their form.  Here, we used the cancellation of collinear divergences order-by-order in $\alpha_s$ to find an appropriate $F(\zg)$. We can also extend \Eq{eq:ff1} beyond the collinear limit by considering full real and virtual matrix elements, leading to finite ${\cal O}(\alpha_s)$ corrections to $p(\zg)$.

As alluded to above, $F_\text{UV}(\zg)$ also has the interpretation of being a UV fixed point from RG evolution.  The collinear divergence of \Eq{eq:ff1} can be absorbed into a renormalized FF, $F^\text{(ren)}(\zg;\mu)$, at the price of introducing explicit dependence on the $\overline{\text{MS}}$ renormalization scale $\mu$. Requiring \Eq{eq:ff1} to be independent of $\mu$ through ${\cal O}(\alpha_s)$ results in the following RG equation for $F^{\text{(ren)}}(\zg;\mu)$:
\begin{align}
\label{eq:RGflow}
\mu\frac{\partial}{\partial \mu}F^\text{(ren)}(\zg;\mu) &= \frac{\alpha_s C_i}{\pi} \\
&
\hspace{-2.5cm}
\times \Biggl( \overline{P}_i(\zg) \Theta(\zg-\zcut) - F^{\text{(ren)}}(\zg;\mu) \int_{z_\text{cut}}^{1/2}dz\,\overline P_i(z)\Biggr)\,.  \nonumber
\end{align}
As $\mu$ goes to $+\infty$, the IR boundary condition is suppressed and $F^\text{(ren)}(\zg;\mu)$ asymptotes to $F_\text{UV}(\zg)$.  

  \begin{figure}[t]
 \includegraphics[height=5cm]{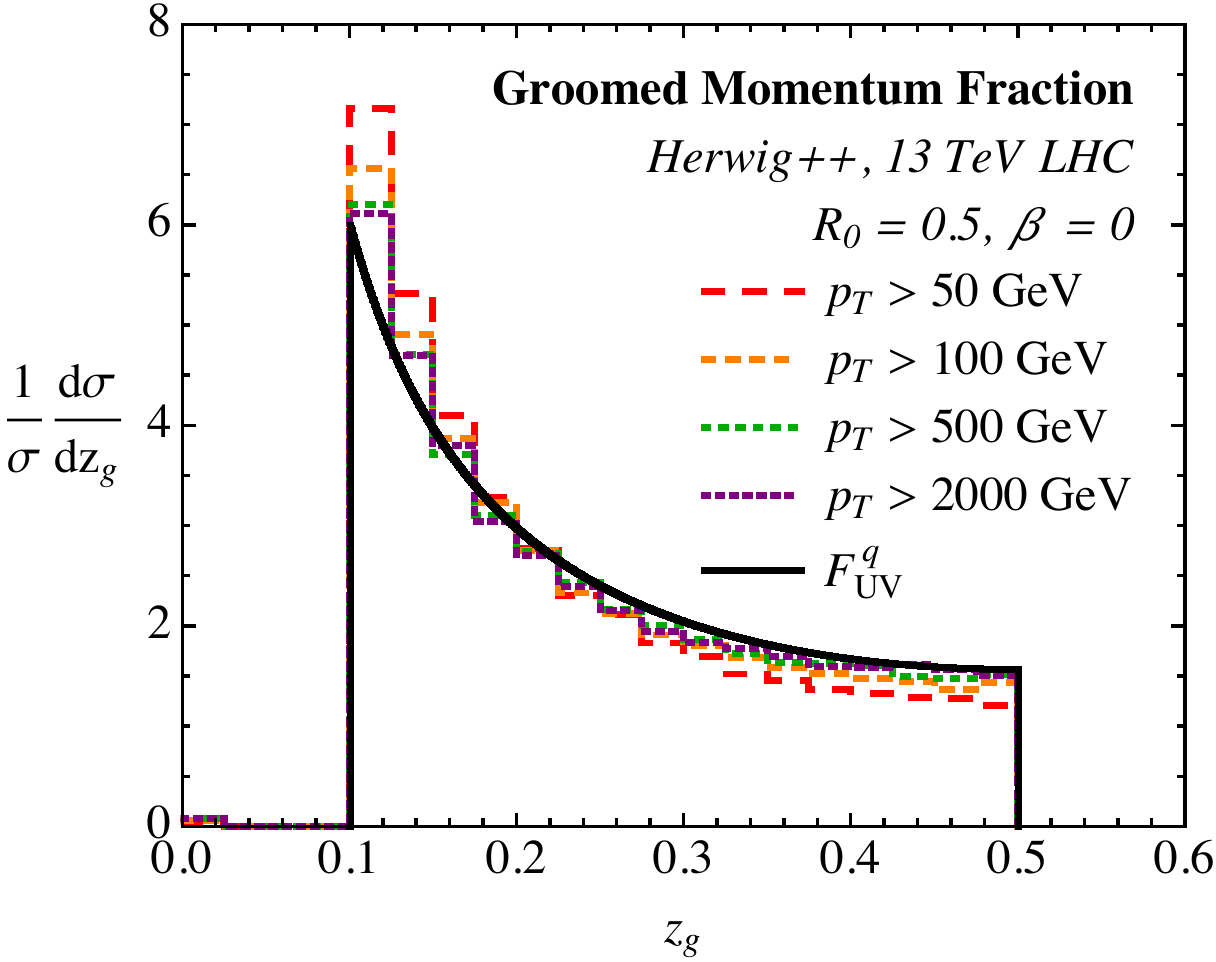}
 \caption{Distributions of $\zg$ for $\beta=0$ and $z_\text{cut}=0.1$ at the 13 TeV LHC, as simulated by \herwigpp 2.6.3. The $p_T$ of the jets ranges from 50 GeV to 2 TeV, and the asymptotic distribution for quark jets, $F_\text{UV}^{q}$ in \Eq{eq:f_uv}, is solid black.
 \label{fig:zf_dists_her}}
 \end{figure}

This UV asymptotic behavior can be tested using parton shower Monte Carlo generators.  In \Fig{fig:zf_dists_her} we show the $\zg$ distribution for $\beta = 0$ for \herwigpp~2.6.3~\cite{Bahr:2008pv} at the 13 TeV LHC, using \fastjet{3.1} \cite{Cacciari:2011ma} and the \textsc{RecursiveTools} contrib \cite{fjcontrib}.  As shown in the supplement, other generators give similar results.   As the jet $p_T$ increases, $p(z_g)$ asymptotes to the form in \Eq{eq:f_uv} (which happens to be nearly identical for quark and gluon jets).  This is due both to the RG flow in \Eq{eq:RGflow}, which suppresses non-perturbative corrections, and the decrease of $\alpha_s$ with energy, which suppresses $\mathcal{O}(\alpha_s)$ corrections to $p(z_g)$. 

In this paper, we gave a concrete definition of Sudakov safety, which extends the reach of pQCD beyond the traditional domain of IRC safe observables.  Even at lowest perturbative order, the $z_g$ example highlights the different analytic structures possible in the Sudakov safe regime, and the FF approach to the IRC safe/unsafe boundary yields new insights into the structure of perturbative singularities.  In addition to being an interesting conceptual result in perturbative field theory, \Eq{eq:sudsafedef} offers a concrete prescription for how to leverage the growing catalog of high-accuracy pQCD calculations (both fixed-order and resummed) to make predictions in the IRC unsafe regime.  This can be done without have to rely (solely) on non-perturbative modeling, enhancing the prospects for precision jet physics in the LHC era.

\begin{acknowledgments}
We thank Gregory Soyez for collaborating in the early stages of this work and Bob Jaffe, Ian Moult, Duff Neill, Michael Peskin, Gavin Salam, George Sterman, and Iain Stewart for enlightening discussions.   This work was supported by the U.S. Department of Energy (DOE) under cooperative research agreement DE-FG02-05ER-41360. J.T.\ is also supported by the DOE Early Career research program DE-FG02-11ER-41741 and by a Sloan Research Fellowship from the Alfred P.\ Sloan Foundation.  S.M.\ is supported by the U.S.\ National Science Foundation, under grant PHY--0969510, the LHC Theory Initiative.
\end{acknowledgments}

\bibliography{sudakov}

\setcounter{equation}{0}

\renewcommand{\theequation}{S-\arabic{equation}}
\renewcommand{\thefigure}{S\arabic{figure}}


\setcounter{page}{1}
\pagenumbering{roman}


\section{Supplemental Material}

The supplemental material contains five additional calculations and analyses to highlight the behavior of Sudakov safe observables, especially at higher orders.

\subsection{Calculating Ratio Observables}

When calculating the ratio observables in \Ref{Larkoski:2013paa}, there is a subtle difference between \Eq{eq:ratio_doublediff} and \Eq{eq:ratio_conditional} with respect to resummation.  In \Refs{Larkoski:2013paa,Larkoski:2014tva} which followed \Eq{eq:ratio_doublediff}, a Sudakov form factor for the joint probability distribution $p(a,b)$ was found, which resums (some) logs of $a$, $b$, and $a/b$.  In contrast, \Eq{eq:ratio_conditional} suggests first resuming logs of $b$ in $p(b)$, and then (optionally) resuming logs of $r$ in $p(r|b)$.  These give slightly different expressions, as we show here.

As in \Refs{Larkoski:2013paa,Larkoski:2014tva}, the observables $a$ and $b$ are recoil-free angularities measured on a jet, which are defined as
\begin{equation}
e_\alpha = \frac{1}{p_{TJ}}\sum_{i\in J} p_{Ti}\left(\frac{R_{i\hat b}}{R_0}\right)^\alpha \,,
\end{equation}
for angular exponent $\alpha> 0$ and jet radius $R_0$.  $R_{i\hat b}$ is the distance in the pseudorapidity-azimuth plane between particle $i$ and the recoil-free broadening axis $\hat{b}$ \cite{Larkoski:2014uqa}.

For two angularities $e_\alpha$ and $e_\beta$, with $\alpha>\beta$, the joint cumulative probability distribution at leading logarithmic accuracy with fixed coupling is \cite{Larkoski:2013paa}
\begin{equation}
\Sigma(e_\alpha,e_\beta)=\exp\left[
-\frac{\alpha_s}{\pi}C_i\left(
\frac{\log^2 e_\beta}{\beta}+\frac{\log^2\frac{e_\alpha}{e_\beta}}{\alpha-\beta}
\right)
\right] \,,
\end{equation}
where $C_i$ is the color factor.  The double differential cross section/joint probability distribution is then
\begin{equation}
p(e_\alpha,e_\beta)=\frac{1}{\sigma}\frac{\df ^2\sigma}{\df e_\alpha\,\df e_\beta} = \frac{\partial^2}{\partial e_\alpha \, \partial e_\beta} \Sigma(e_\alpha,e_\beta) \,.
\end{equation}
To determine the distribution for the ratio $r=e_\alpha/e_\beta$, we can then insert this into \Eq{eq:ratio_doublediff}.  The full expression is given in \Ref{Larkoski:2013paa}, and the lowest order terms in the small $\alpha_s$ expansion are
\begin{align}\label{eq:ratio_ll}
p(r)&=\frac{1}{\sigma}\frac{\df \sigma}{\df r}=\frac{1}{\sigma}\int \df e_\alpha \df e_\beta \frac{\df^2 \sigma}{\df e_\alpha\, \df e_\beta} \, \delta\left(
r-\frac{e_\alpha}{e_\beta}
\right) \\
&= \sqrt{\alpha_s}\frac{\sqrt{\beta C_i}}{\alpha-\beta}\frac{1}{r} -2\frac{\alpha_s}{\pi} \frac{C_i (\alpha-2\beta)}{(\alpha-\beta)^2}\frac{\log r}{r}+{\cal O}(\alpha_s^{3/2}) \,. \nonumber
\end{align}
This distribution can be considered the leading logarithmic distribution for $r$ because it was calculated from the double differential cross section calculated to leading logarithmic accuracy in $a$, $b$, and $a/b$.

Alternatively, we can calculate $p(r)$ as in \Eq{eq:ratio_conditional}, where the conditional probability, calculated to fixed order, is integrated against the resummed distribution for the denominator.  To leading logarithmic accuracy with fixed coupling, the cumulative distribution for $e_\beta$ is
\begin{equation}
\Sigma(e_\beta) = \exp\left[
-\frac{\alpha_s}{\pi}\frac{C_i}{\beta}\log^2 e_\beta
\right] \,.
\end{equation}
The probability distribution of $e_\beta$ is thus
\begin{equation}
p(e_\beta) = \frac{1}{\sigma}\frac{\df \sigma}{\df e_\beta} = \frac{\partial}{\partial e_\beta} \Sigma(e_\beta) \,.
\end{equation}
Calculating the conditional probability using the most singular terms in the splitting function, we find 
\begin{equation}
p(r|e_\beta) = \frac{\beta}{\alpha-\beta}\frac{1/r}{\log 1/e_\beta} \Theta(1-r)\Theta\left(
r-e_\beta^{\frac{\alpha-\beta}{\beta}}
\right) \,
\end{equation}
to lowest order, which is indeed a normalized conditional probability distribution.  Using the method of \Eq{eq:ratio_conditional} yields the probability distribution for the ratio $r$:
\begin{align}
p(r) &= \int \df e_\beta\, p(e_\beta)\, p(r|e_\beta)\\
&= \sqrt{\alpha_s}\frac{\sqrt{\beta C_i}}{\alpha-\beta}\frac{1}{r}+2\frac{\alpha_s}{\pi} \frac{C_i \beta}{(\alpha-\beta)^2}\frac{\log r}{r}+{\cal O}(\alpha_s^{3/2}) \,. \nonumber
\end{align}
The term at ${\cal O}(\sqrt{\alpha_s})$ agrees with \Eq{eq:ratio_ll}, but the term at ${\cal O}(\alpha_s)$, and higher terms, generically do not.  This emphasizes that to determine the formal accuracy of a Sudakov safe observable requires specifying the accuracy of all components in its calculation.

\subsection{Heuristic for the $\as$ Expansion}

In \Tab{Tab:behavior}, we drew a link between the singularities present in $z_g$ and the expected $\as$ expansion for $p(z_g)$.  Here, we give a heuristic way to understand this behavior.

For a generic resummed observable $s\in (0,1)$, the probability distribution $p(s)$ can be written as
\begin{equation}
p(s) = \frac{\df}{\df s} e^{f(s)} = f'(s) \, e^{f(s)},
\end{equation}
where $f(s)$ is some function, and $\Sigma(s) = e^{f(s)}$ is the resummed cumulative distribution for $s$.  Depending on whether $s$ has both soft and collinear singularities or just collinear ones, $f(s)$ is expected to take different forms:
\begin{align}
f_{\rm sc} (s) &= \sum_{k = 1}^{\infty} c_k \, \as^k \log^{k+1} s + \ldots,  \label{eq:f_sc} \\
f_{\rm c} (s) &= \sum_{k = 1}^{\infty} d_k \, \as^k \log^{k} s + \ldots,  \label{eq:f_c}
\end{align}
where the ellipses ($\ldots$) stand for terms with additional $\as$ suppression.  We say that $f_{\rm sc} (s)$ has ``double log'' behavior (since the lowest term is $\as \log^2 s$), while $f_{\rm c} (s)$ is ``single log'' (for $\as \log s$).

The functional form of $p(u|s)$ is not needed to derive our heuristic, though a few facts about $p(u|s)$ are important.  First, since $s$ regulates the divergences in $u$, $p(u|s)$ must have a valid Taylor expansion in $\as$.  Second, because $s$ itself has singularities, $p(u|s)$ will necessarily have dependence on $\log s$ related to the structure of $p(s)$.  Third, because the conditional probability distribution is normalized as $\int \df u \, p(s|u) = 1$, $p(u|s)$ has to start at $\mathcal{O}(\as^0)$.  In particular, at lowest order 
\begin{align} 
p_{\rm sc}(u|s) &= \frac{g(u,s)}{\log s} + \mathcal{O}(\as), \\
p_{\rm c}(u|s) &= h(u,s) + \mathcal{O}(\as), 
\end{align}
where we have pulled out an extra $\log s$ factor in the soft/collinear case, such that $g(u,s)$ and $h(u,s)$ only have power-suppressed dependence on $s$.

From these generic forms for $p(s)$ and $p(u|s)$, we can determine $p(u)$ using \Eq{eq:sudsafeone},
\begin{align}
p(u) &= \int \df s \, f'(s) \, e^{f(s)} \, p(u|s) \\
& = \int \df x \, e^{x} \, p(u|s(x)), \label{eq:puasxint}
\end{align}
where we have introduced the change of variables
\begin{equation}
x = f(s), \qquad \df x = f'(x) \, \df s,
\end{equation}
and we expect $x \in (-\infty,0)$.  With this change of variables, the $\as$ dependence of $p(u)$ resides entirely in $p(u|s(x))$.  In general, there is no closed form for $s(x)$, but we can determine it order by order in $\as$ by inverting the series in \Eqs{eq:f_sc}{eq:f_c}.  Depending on the divergence structure of $s$, there are different forms:
\begin{align}
\text{sc}: \quad \as \log s(x) &= \sqrt{\frac{\as x}{c_1}} + \mathcal{O}(\as), \\
\text{c}: \quad \as \log s(x) &= \frac{x}{d_1} + \mathcal{O}(\as).
\end{align}
In the soft/collinear case, the $\sqrt{\as}$ factors in $\log s$ mean that $p(u|s(x))$ will have a $\sqrt{\as}$ expansion when expressed as a function of $x$.  By contrast, in the collinear only case, $p(u|s(x))$ will still have an ordinary $\as$ expansion.

The last ingredient is the starting order of the expansion.  Plugging in the lowest order expressions for $p(u|s)$, we can evaluate $p(u)$ using \Eq{eq:puasxint}, up to power corrections:
\begin{align}
p_{\rm sc} (u) &= \sqrt{- \alpha_s c_1 \pi} \, g(u,0) + \mathcal{O}(\as), \\
p_{\rm c} (u) &= h(u,0)  + \mathcal{O}(\as).
\end{align}
This confirms the expected expansions in \Tab{Tab:behavior}.  Note that in the collinear only case, there is no leading dependence on $d_1$, though $d_1$ will show up at $\mathcal{O}(\as)$.

\subsection{The $\zg$ Distribution at Higher Accuracy}

Via \Eq{eq:zfeq}, we calculated $p(\zg)$ to the lowest non-trivial order. Here, we discuss how to improve the accuracy of this calculation through running coupling effects. The following discussion is valid for any $\beta$.

We start with the groomed radius distribution $p(\rg)$, for which the all-orders resummation was derived in~\Ref{Larkoski:2014wba}. Including running coupling effects in \Eq{eq:resum_r}, we obtain
\begin{equation} \label{eq:resum_r_rc}
p(r_g) =  \frac{\df}{\df \rg} e^{f(r_g)},
\end{equation}
where 
\begin{align} 
\hspace{-0.2cm} f(r_g) &=  -\frac{2  C_i}{\pi} \int_{\rg}^1\frac{\df \theta}{\theta}\int_0^1 \df z \, P_i(z) \,\as\left(\tilde{z} \, \theta \, p_T R_0\right) \Theta_\text{cut} \nonumber\\
&= -\frac{2  C_i}{\pi} \int_{\rg}^1\frac{\df \theta}{\theta}\int_{\zcut 
\theta^\beta}^{1/2} \!  \df z \, \overline{P}_i(z) \,\as\left(z \, \theta \, p_T R_0\right) ,
\end{align}
$\tilde{z}=\min(z,1-z)$, and $\Theta_\text{cut}$ is defined in \Eq{eq:theta_cut}.  When summed over final states, the quark and gluon splitting functions are
\begin{align}
P_q(z)&=\frac{1+(1-z)^2}{2z},  \label{splittings_q} \\
\hspace{-0.25cm}
 P_g(z)&=
\frac{1-z}{z}+\frac{z(1-z)}{2}\nonumber \\
&\qquad
+\frac{n_f T_R}{2C_A}\left[
z^2+(1-z)^2
\right] . \label{splittings_g}
\end{align}
Exploiting $z\leftrightarrow(1-z)$ symmetry, we have written the gluon splitting function in such a way that it exhibits a singularity only when $z\to 0$.
As noted in~\Ref{Larkoski:2014wba},  \Eq{eq:resum_r_rc} is accurate to single-logarithmic accuracy, provided that $\as$ is evaluated in the CMW scheme~\cite{Catani:1990rr}. This is because $\rg$ is set by just one splitting, with no multiple-emissions contribution. The expression in \Eq{eq:resum_r_rc} only differs from the corresponding one in~\Ref{Larkoski:2014wba} because of the more sophisticated $\Theta_\text{cut}$ treatment, which takes into account finite $z$ corrections.

Next, we address the conditional probability $p(\zg | \rg)$. There are various strategies to compute this quantity.  For example, we could use exact fixed-order matrix elements or we could embark on a systematic all-orders calculation.  Here, we will show an intermediate approach, working in the collinear limit for the matrix element, but including a tower of all-orders contributions originating from $\as$ running.  Specifically, we write
\begin{equation}
p(\zg | \rg) = \frac{p^*(\zg,\rg)}{p^*(\rg)} \, \Theta(\zg-\zcut \rg^\beta)\, ,
\end{equation}
where
\begin{align}
p^*(\zg , \rg) &= \frac{2 C_i}{\pi}
  \int_0^1 \frac{\df \theta}{\theta} \int_{0}^1 \df z \, P_i (z)\, \Theta_\text{cut}   \, \delta(\tilde{z}-\zg)  \nonumber\\
  & \quad \times\delta(\theta-\rg)  \, \as\left(\tilde{z} \, \theta \, p_T R_0\right) 
  \nonumber\\&
  =  \frac{2 C_i}{\pi r_g} \as\left(\zg \, \rg \, p_T R_0\right) \overline{P}_i(\zg)
\end{align}
has $\alpha_s$ evaluated at the proper scale of the emission,
and 
\begin{align}
p^*(\rg) &= \int_{\zcut \rg^\beta}^{1/2} \df \zg \, p^*(\zg , \rg) \nonumber\\
&= \frac{2 C_i}{\pi \rg} \int_{z_\text{cut}\rg^\beta}^{1/2} \df \zg \, \overline{P}_i(\zg)\,
   \as\left(\zg \, \rg \, p_T R_0\right)
\end{align}
ensures that $p(\zg | \rg)$ is properly normalized.

 \begin{figure}[t]
 \includegraphics[height=5cm]{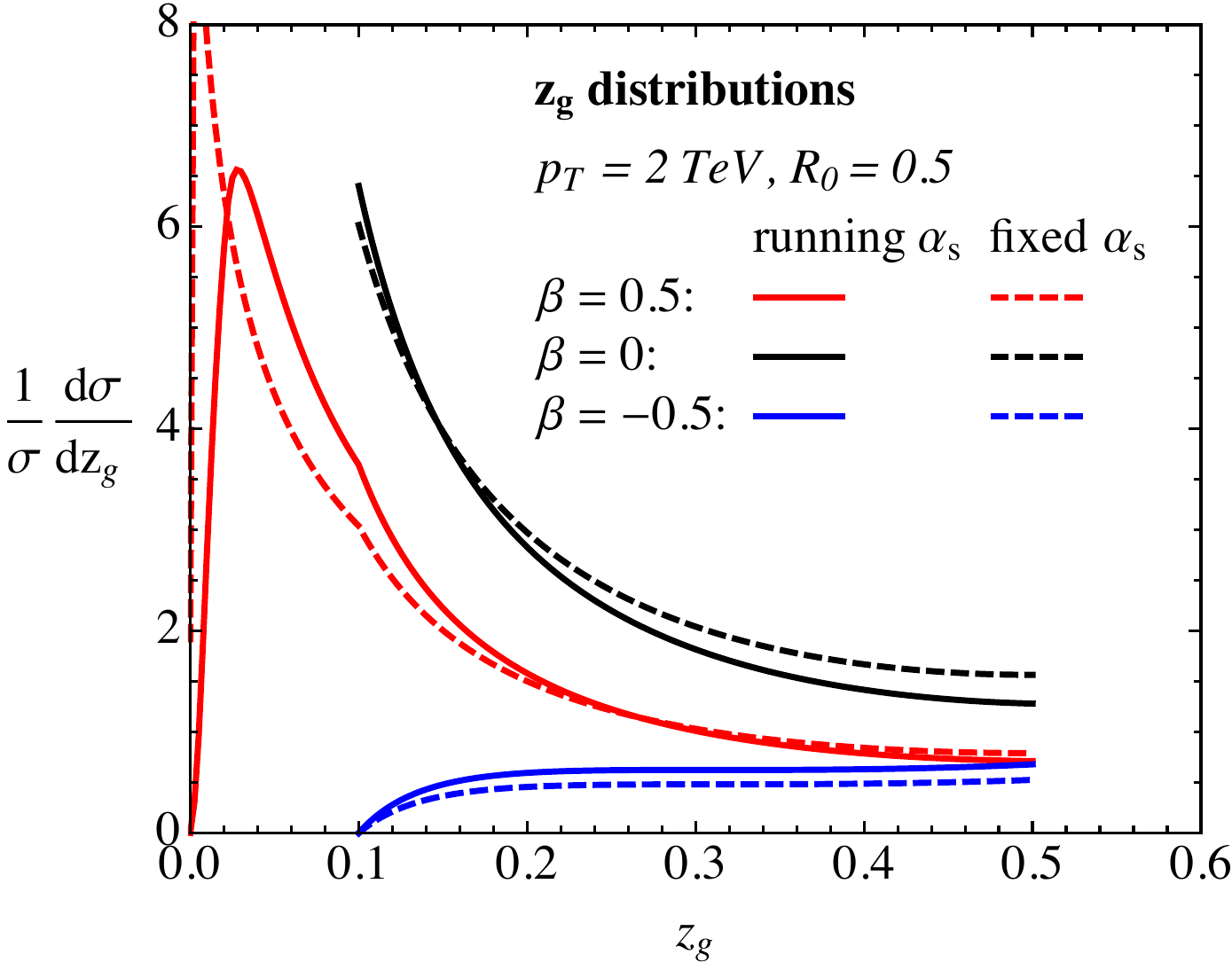}
 \caption{Resummed distributions of $z_g$ for $\zcut=0.1$ and three different values of $\beta=0.5$, $0$, $-0.5$.  We take $p_T=2$~TeV and $R_0=0.5$.  Solid curves represent resummation with running coupling, while dashed ones are evaluated at fixed coupling $\as(p_T R_0)=0.087$.
 \label{fig:zf_dists_rc}}
 \end{figure}

With these running coupling improvements, $p(\zg)$ can be computed using the Sudakov safe definition in \Eq{eq:zfeq}. To regularize the Landau pole in $\as$, we freeze its running at the non-perturbative scale $\mu_\text{NP}=1$~GeV.  Our results are shown in \Fig{fig:zf_dists_rc} and compared to the fixed coupling case with $\as(p_T R_0)=0.087$.

\subsection{Renormalization Group Evolution of the Fragmentation Function}

As discussed in the paper, the collinear divergence of \Eq{eq:ff1} can be absorbed into a renormalized FF, at the price of introducing explicit scale dependence.  In dimensional regularization in the $\overline{\text{MS}}$ scheme, the renormalized $F(\zg)$ to ${\cal O}(\alpha_s)$ is 
\begin{align} \label{eq:Fren}
F^{\text{(ren)}}(\zg;\mu)&=F(\zg) -\left( \frac{1}{2\epsilon} +c  \right) \frac{\alpha_s C_i}{\pi} \\
&
\hspace{-1cm}
\times 
\left( \overline{P}_i(\zg)  \Theta(\zg-\zcut) - F(\zg) \int_{z_\text{cut}}^{1/2}dz\, \overline{P}_i(z)\right)\,,\nonumber
\end{align}
where $\epsilon$ is the dimensional regularization parameter, $c=\log (4\pi e^{- \gamma_E})$, and $\mu$ is the renormalization scale.   
Requiring the cross section to be independent of $\mu$ through ${\cal O}(\alpha_s)$ results in an RG equation for $F^\text{(ren)}(\zg;\mu)$:
\begin{align}\label{eq:rg_f}
\mu\frac{\partial}{\partial \mu}F^\text{(ren)}(\zg;\mu) &= \frac{\alpha_s(\mu) }{\pi}C_i \\
&
\hspace{-2.5cm}
\times \Biggl( \overline{P}_i(\zg) \Theta(\zg-\zcut) - F^{\text{(ren)}}(\zg;\mu) \int_{z_\text{cut}}^{1/2}\df z\,\overline P_i(z)\Biggr)\,,   \nonumber
\end{align}
up to corrections at ${\cal O}(\alpha_s^2)$.  Compared to \Eq{eq:RGflow}, here we have explicitly introduced the $\mu$ dependence in $\alpha_s$.  

To solve for $F^{\text{(ren)}}(\zg;\mu)$, we first find the solution to the homogeneous equation:
\begin{align}
&\mu\frac{\partial}{\partial \mu}F_h^\text{(ren)}(\zg;\mu) \\
&\qquad= -\left[\frac{\alpha_s(\mu) C_i }{\pi} \int_{z_\text{cut}}^{1/2} \df z\,\overline P_i(z)\right] F_h^{\text{(ren)}}(\zg;\mu)\,.  \nonumber
\end{align}
The solution is
\begin{align}
&F_h^\text{(ren)}(\zg;\mu) \\
&\quad= F_0(z_g;\mu_0)\exp\left[
- \frac{C_i}{\pi}\int_{z_\text{cut}}^{1/2}dz\,\overline P_i(z) \int_{\mu_0}^{\mu} \frac{\df \mu'}{\mu'} \alpha_s(\mu')
\right] \,,\nonumber
\end{align}
where $F_0(z_g;\mu_0)$ is the boundary value defined at the infrared scale $\mu_0$.  Using the definition of the $\beta$-function,
\begin{equation}
\mu\frac{\partial \alpha_s}{\partial \mu} = \beta(\alpha_s)\,,
\end{equation}
the integral over the scale $\mu'$ can be exchanged for an integral over $\alpha_s$ itself:
\begin{equation}
 \int_{\mu_0}^{\mu} \frac{\df \mu'}{\mu'} \alpha_s(\mu') \Rightarrow \int_{\alpha_s(\mu_0)}^{\alpha_s(\mu)} \frac{ \df \alpha'}{\beta(\alpha')}\alpha'.
\end{equation}
The one loop $\beta$-function is
\begin{equation}
\beta(\alpha_s)=-\alpha_s^2 \beta_0\,,
\end{equation}
which then produces the homogeneous solution
\begin{equation}
F_h^\text{(ren)}(\zg;\mu)=F_0(z_g;\mu_0) \left(\frac{\alpha_s(\mu)}{\alpha_s(\mu_0)} \right)^{\frac{C_i}{\pi \beta_0}\int_{z_\text{cut}}^{1/2} \df z\,\overline P_i(z)} \,.
\end{equation}
The exponent is found by integrating the quark and gluon splitting functions in  \Eqs{splittings_q}{splittings_g}:
\begin{align}
\int_{z_\text{cut}}^{1/2} \df z\,\overline P_q(z) &=\log
   \left(\frac{1-z_{\text{cut}}}{z_{\text{cut}}}\right) -\frac{3}{
   4} + \frac{3}{2} z_{\text{cut}}\,, \\
\int_{z_\text{cut}}^{1/2} \df z\,\overline P_g(z)&=  \log
   \left(\frac{1-z_{\text{cut}}}{z_{\text{cut}}}\right) + \left(\frac{n_f}{6 C_A} -\frac{11}
   {12} \right) \nonumber \\
& \quad +2 z_{\text{cut}}
   \left(1-\frac{n_f}{4C_A}\right) -\frac{z_{\text{cut}}^2}{2}
   \left(1-\frac{n_f}{C_A}\right)  \nonumber \\
   & \quad + \frac{z_{\text{cut}}^3}{3} 
   \left(1-\frac{n_f}{C_A}\right).
   \end{align}

Now, we must find a particular solution to \Eq{eq:rg_f}.  The simplest approach is to assume that $F^\text{(ren)}(\zg;\mu)$ is independent of $\mu$, which requires:
\begin{align}
F_p^\text{(ren)}(\zg;\mu)= \frac{\overline{P}_i(\zg) }{\int_{z_\text{cut}}^{1/2} \df z\,\overline P_i(z)}\Theta(\zg-\zcut) \,.
\end{align}
The full solution to \Eq{eq:rg_f} is then the sum of the homogeneous and particular solutions:
\begin{align}
F^\text{(ren)}(\zg;\mu) &= F_h^\text{(ren)}(\zg;\mu)+F_p^\text{(ren)}(\zg;\mu) \nonumber \\
& =F_0(z_g;\mu_0) \left(\frac{\alpha_s(\mu)}{\alpha_s(\mu_0)} \right)^{\frac{C_i}{\pi \beta_0}\int_{z_\text{cut}}^{1/2} \df z\,\overline P_i(z)}  \nonumber\\
&\quad+ \frac{\overline{P}_i(\zg) }{\int_{z_\text{cut}}^{1/2} \df z\,\overline P_i(z)}\Theta(\zg-\zcut) \,. \label{eq:rge_running_full}
\end{align}
By asymptotic freedom of $\alpha_s$, as $\mu\to \infty$, the homogeneous solution is suppressed and any dependence on the infrared boundary condition $F_0(z_g;\mu_0)$ vanishes.  Therefore, the asymptotic distribution is
\begin{align}
F^\text{(ren)}(\zg;\mu\to \infty)= \frac{\overline{P}_i(\zg) }{\int_{z_\text{cut}}^{1/2}\df z\,\overline P_i(z)}\Theta(\zg-\zcut) \,.
\end{align}
This agrees with the Sudakov safe calculation in \Eq{eq:beta_zero_pzg}, and makes a definite prediction for the high energy behavior of this observable.

We can also use \Eq{eq:rge_running_full} to estimate the scaling of non-perturbative corrections to $p(z_g)$.  The boundary condition $F_0(z_g;\mu_0)$ is $\mathcal{O}(1)$ and fully non-perturbative, but in the small $\zcut$ limit it scales away approximately as
\begin{equation}
\label{eq:np_scaling}
\left(1 + \as(\mu_0) \beta_0 \log \frac{\mu}{\mu_0} \right)^{- \frac{C_i}{\pi \beta_0} \log \frac{1}{\zcut}}.
\end{equation}
Though not quite a power law suppression, it would be a power law in the $\beta_0 \to 0$ limit
\begin{equation}
\left(\frac{\mu_0}{\mu}\right)^{\frac{\alpha_s C_i}{\pi} \log \frac{1}{\zcut}} + \mathcal{O}(\beta_0) \, ,
\end{equation}
with typical exponent $\simeq 0.1$ ($\simeq 0.2$) in the quark (gluon) case.  In general, because of the larger $C_i$, we predict that gluon-initiated jets will saturate $F_{\rm UV}(z_g)$ faster than quark-initiated jets.  Beyond $F(z_g)$, there may be shape function corrections as with IRC safe observables.

Finally, we have found evidence that the FF UV fixed point in \Eq{eq:f_uv} may be one-loop exact.  We have verified this explicitly at ${\cal O}(\alpha_s^2)$ in the strongly-ordered collinear limit but have not yet attempted a complete proof.  We observe that the collinear singularity in $z_g$ is resolved at any non-zero value of $r_g$.   We therefore expect that the only kinds of singularities present are those associated with a $1 \to 2$ splitting, which are already included in \Eq{eq:rg_f}.  No additional collinear singularities are expected to appear in $1 \to 3$ splittings, since either the third parton generates a finite value of $r_g$ (in which case all singularities are regulated), the third parton fails the soft drop condition (in which case the kinematics reverts to $1 \to 2$), or the third parton is collinear with the other two (in which case the real $1 \to 3$ collinear singularity should cancel against the virtual $1 \to 2$ one).

\subsection{Monte Carlo Analysis of $z_g$ for $\beta = 0$}

When comparing our analytic calculation of $p(\zg)$ at $\beta=0$ to Monte Carlo generators in \Fig{fig:zf_dists_her}, we only showed results for \herwigpp~2.6.3 \cite{Bahr:2008pv}.  Here, we include two additional generators:  \pythia~8.201~\cite{Sjostrand:2007gs} and \sherpa~2.1.1~\cite{Gleisberg:2008ta}. Although not reported here, we checked that the features below are also observed in the \vincia~1.1.3 \cite{Giele:2007di} antenna shower applied to high energy electron-positron collisions.  In all cases, we used \fastjet 3.1 \cite{Cacciari:2011ma} to reconstruct jets and the \textsc{RecursiveTools} contrib \cite{fjcontrib} to implement soft drop grooming.

\begin{figure}[h]
 \includegraphics[height=5cm]{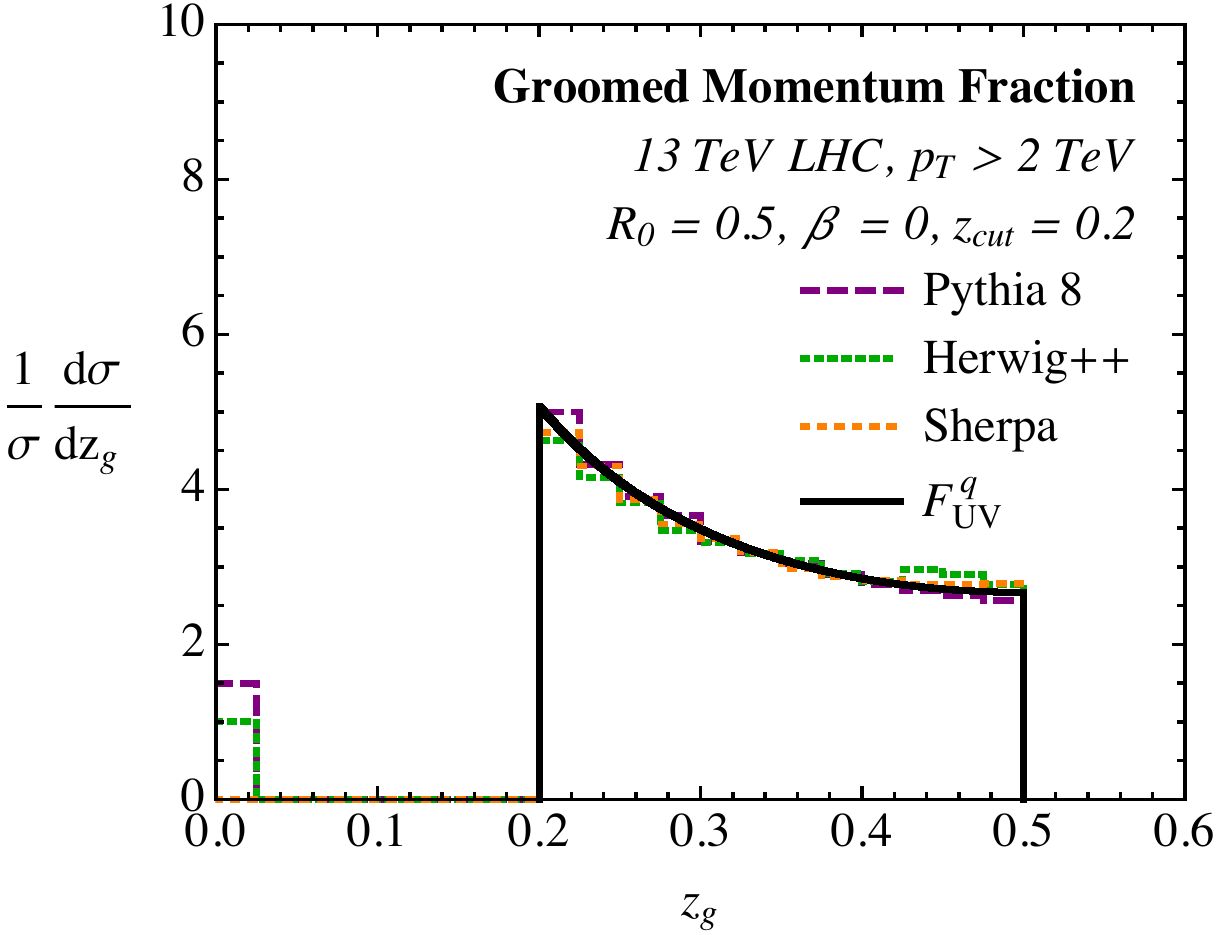} 
\\ \ \\ 
 \includegraphics[height=5cm]{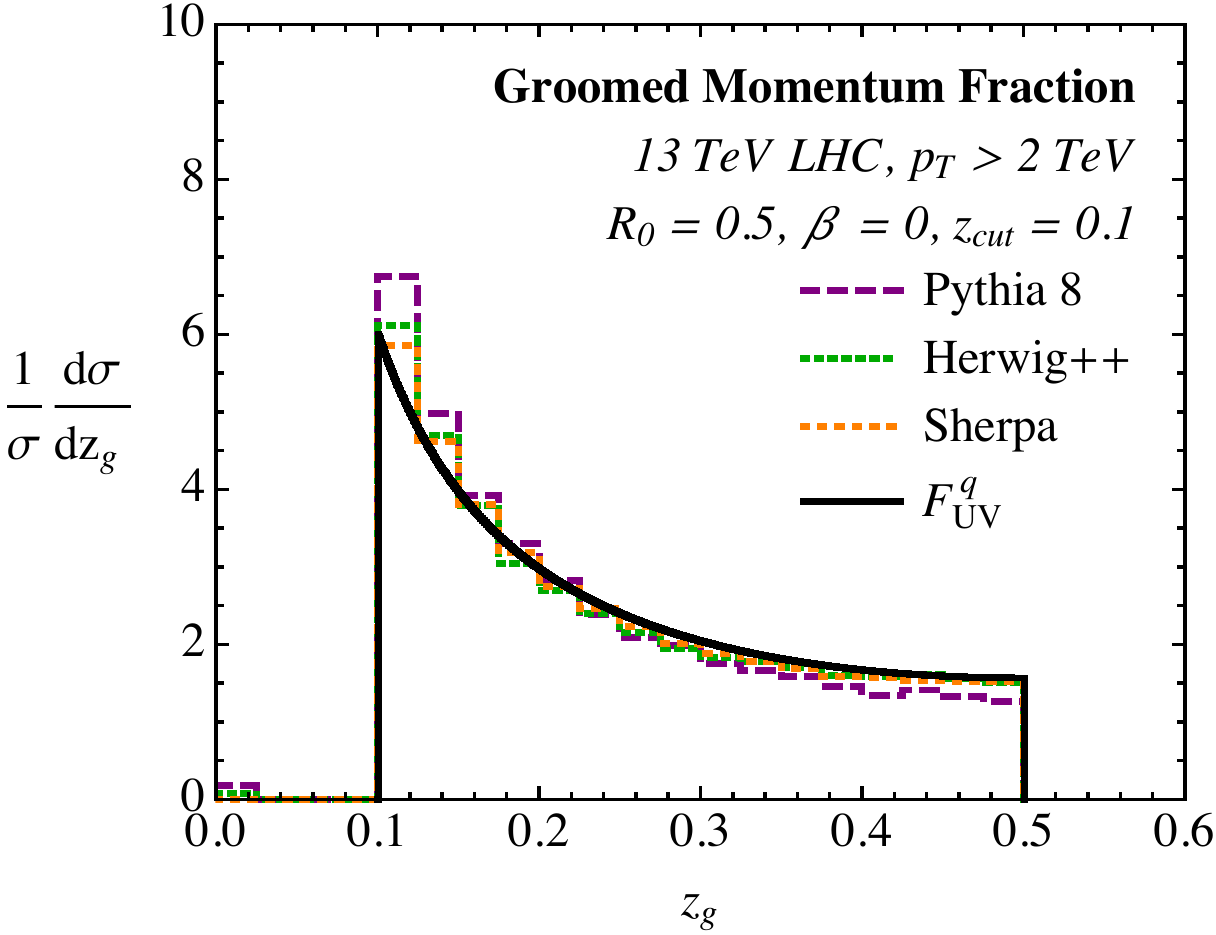}
 \\ \ \\ 
 \includegraphics[height=5cm]{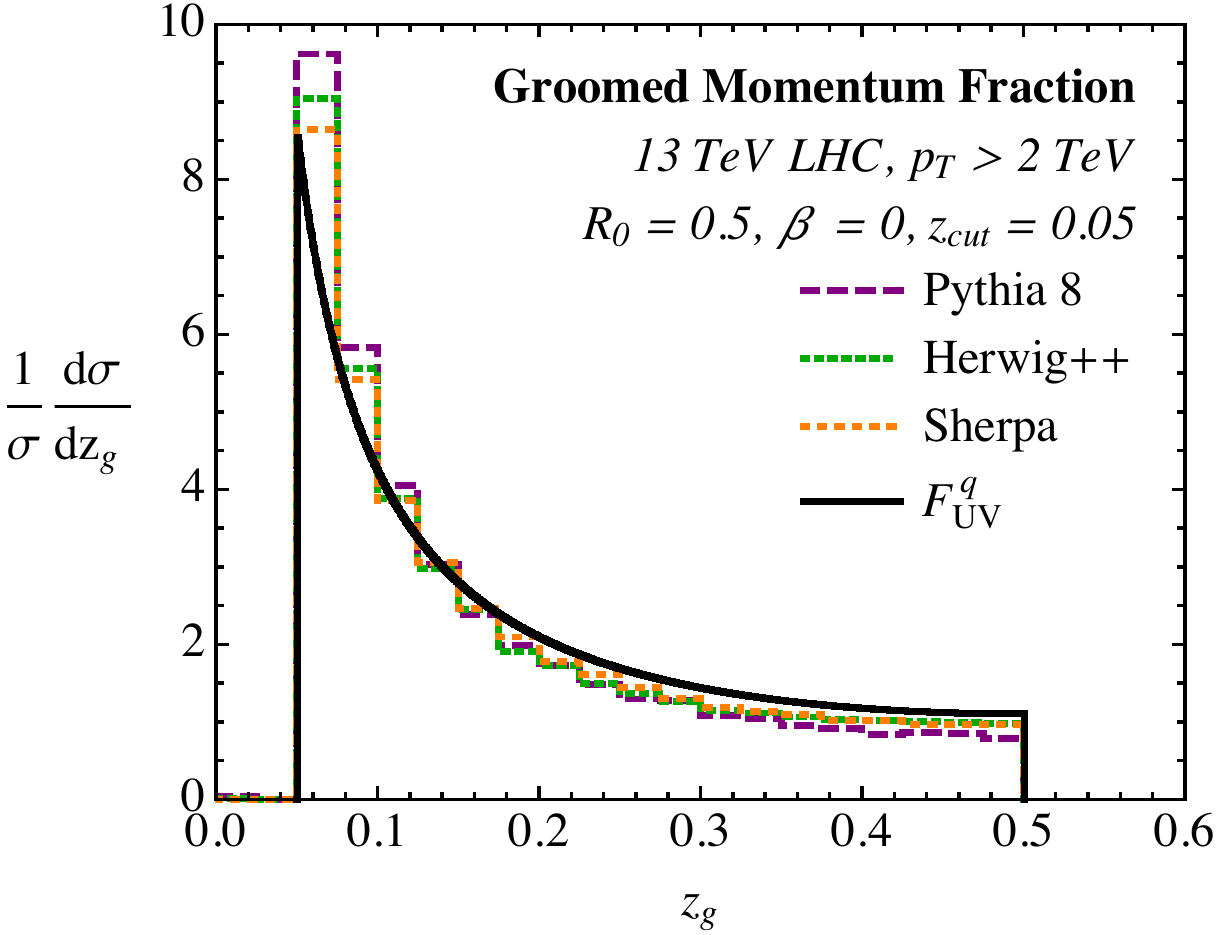}
 \caption{Distributions of $\zg$ for $\beta=0$ at the 13 TeV LHC as simulated from  \herwigpp, \pythia, and \sherpa samples.  From top to bottom, the $\zcut$ values are $0.2$, $0.1$ and $0.05$.  Jets are selected with $p_T>2$~TeV and jet radius $R_0 = 0.5$.
 \label{fig:zf_dists}}
 \end{figure}

In \Fig{fig:zf_dists}, we show the $\zg$ distribution at $\beta=0$ for three different values of the energy cut, $\zcut=0.2$, $0.1$, and $0.05$.  All samples are at the 13 TeV LHC, including hadronization effects as well as the default underlying event models. For simplicity we only show the $\zg$ distributions for jets with large transverse momentum, $p_T>2$~TeV, in order to reduce hadronization corrections.  Jets at $z_g = 0$ correspond to situations where the soft drop groomer gives a one-prong configuration, and those events are not used for normalizing $p(z_g)$.

All distributions are in decent agreement with the fixed-coupling analytic prediction in \Eq{eq:f_uv}.  The agreement is particularly good for larger values of $\zcut$ but degrades as $\zcut$ gets smaller, especially in the \pythia\ sample.  We interpret the primary difference between our analytic calculation and the Monte Carlo generators as arising from running coupling effects (see \Fig{fig:zf_dists_rc}) and from residual dependence on non-perturbative hadronization corrections.   As $\zcut$ decreases, the phase-space for emissions which build up the Sudakov form factor extends more and more into the soft region, eventually picking up contributions from non-perturbative emissions.  This affects small values of $z_g$, which then distorts the overall $p(z_g)$ distribution after normalization.

Though not shown, the small $\zcut$ distortion is more pronounced for quark jets than for gluon jets.  This is expected from the analysis of \Eq{eq:np_scaling}, since the larger gluon color factor means that more of the distribution is described by wide angle perturbative emissions.  Another potential source of distortion is large logarithms of $z_g$ and $\zcut$ which have not been resummed.  Some of these logarithms can be captured through running coupling effects, and preliminary investigations indicate that remaining logarithms may give a small contribution, owing to the single-emission nature of $z_g$.

\end{document}